# Nanometer-scale striped surface terminations on fractured SrTiO$_3$ surfaces

Nathan P. Guisinger,[1*] Tiffany S. Santos,[1] Jeffrey R. Guest,[1] Te-Yu Chien,[1] Anand Bhattacharya,[1] John W. Freeland,[1] Matthias Bode[1]

[1]Argonne National Laboratory, Argonne, IL 60439

e-mail: nguisinger@anl.gov

Corresponding Author's Current Address:

Argonne National Laboratory

Center for Nanoscale Materials

9700 South Cass Avenue, Bldg. 440

Argonne, IL 60439

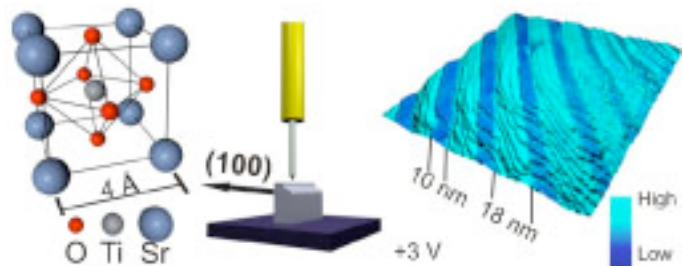

# ABSTRACT

Using cross-sectional scanning tunneling microscopy on *in situ* fractured $SrTiO_3$, one of the most commonly used substrates for the growth of complex oxide thin films and superlattices, atomically smooth terraces have been observed on (001) surfaces. Furthermore, it was discovered that fracturing this material at room temperature results in the formation of stripe patterned domains having characteristic widths (~10 nm to ~20 nm) of alternating surface terminations that extend over a long-range. Spatial characterization utilizing spectroscopy techniques revealed a strong contrast in the electronic structure of the two domains. Combining these results with topographic data, we are able to assign both $TiO_2$ and $SrO$ terminations to their respective domains. The results of this proof-of-principle experiment reveal that fracturing this material leads to reproducibly flat surfaces that can be characterized at the atomic-scale and suggests that this technique can be utilized for the study of technologically relevant complex oxide interfaces.

MANUSCRIPT TEXT

Oxides containing transition elements within the d-block of the periodic table exhibit a wide range of physical properties and structure, while comprising one of the most interesting set of materials currently studied[1]. Striking phenomena has also been observed in the study of oxide thin films, heterostructures, and superlattices, in which some layered oxide structures result in material attributes that are not observed within their individual constituents[2-7]. Many spatially resolved tools for determining chemical structure at an interface exist[8]. Scanning probe techniques have been applied to a wide range of oxide materials[9-14]. However, and particularly with $SrTiO_3$, these techniques have been utilized to investigate the surface morphology and resulting surface reconstructions following various sample preparation techniques that involve chemical and thermal processing[15-19]. Here we show by cross-sectional STM that fracturing $SrTiO_3$ (Fig. 1a), the most frequently used support for oxide thin films and layered oxide structures, at room temperature results in atomically smooth terraces that form striped pattern domains with characteristic widths (~ 10 nm to ~ 30 nm). Scanning tunneling spectroscopy reveals that the terraces consist of alternating surface terminations of low-energy $TiO_2$ domains and higher-energy SrO domains. This application of cross-sectional STM towards the study of complex oxides and ultimately oxide interfaces provides a pristine environment to observe physical phenomena, such as nanometer striped surface terminations, that would otherwise be lost.

Experiments were performed using commercially purchased $Nb:SrTiO_3(100)$ (0.2% Nb doping), which does not have a natural cleavage plane. The backsides of the samples were scribed *ex situ* utilizing a high precision dicing saw, with the scribe depth roughly 75% through the sample. The surface of the sample, the scribe cut, and the desired fracture are all along {100} planes. The scribed $Nb:SrTiO_3(100)$ is mounted vertically to a sample holder and introduced into an ultrahigh vacuum (UHV) system that houses a commercial STM. The samples were fractured *in situ* at room temperature by applying a force to the backside of the sample with the fracture initiating at the scribe. Following this procedure, the samples were loaded into the STM, whose probe was then positioned directly over the fractured region. All characterization of the fractured Nb:SrTiO$_3$(100) was performed at room temperature.

The gradient enhanced STM image (Fig. 1b) illustrates two distinct regions of topography that have been observed on fractured SrTiO$_3$. In the upper right hand portion of this image all of the step-edges run perpendicular to the [100] direction. In this region the steps are progressing from higher topography in the bottom right to lower topography in the top left. There is a transition when progressing to the bottom left-hand portion of the image showing the other characteristic region in which the step-edges are not running perpendicular to the [100]. Although highly stepped, it is clear that fracturing the sample results in regions of atomically flat {001} planes. This observation, which is highly reproducible, illustrates that cross-sectional STM techniques can be applied to conductive 3-dimensional complex oxides. As we will show below, the resulting fracture leads to well-ordered striped domains of alternating surface terminations, whose widths have characteristic lengths on the nanometer scale.

These domains, which can be difficult to detect from the topography alone, become apparent when analyzing the scanning tunneling spectroscopy (STS) data that was concurrently measured with the topography. In this case, *dI/dV* conductance maps were acquired at a given energy (i.e. constant sample bias) via lock-in detection. These conductance maps can be interpreted as spatially resolved values that are proportional to the local density of states (LDOS) for a given energy. When this data is overlayed onto the topographic data (Fig. 1c, imaged at +3 V sample bias) the striped pattern becomes apparent and shows two strongly contrasting regions in the LDOS.

The striped domains exhibit long-range order over the surface and appear to have two characteristic lengths that are on average 12 nm (dark stripes representing lower sample LDOS) for one domain and 24 nm (bright stripes with enhanced LDOS) for the other. It is important to note that these domains are unaffected by the direction or change in direction of the step-edges, as illustrated in Fig. 1c. Previously reported bulk measurements and theory indicate that fractured SrTiO$_3$ is comprised of a mixture of $TiO_2$ and SrO surface terminations[20-22]. However, to date there are no direct spatial observations of how the mixed termination occurs. Therefore, it is not unreasonable to speculate that these striped patterns, which exhibit a strong electronic contrast, are in fact the two different surface terminations. While STM is not capable of direct chemical identification, STS can be utilized to shed light on the origin and chemical composition of these domains.

Closer inspection of the topographic STM data, particularly in regions containing step-edges that run perpendicular to the [100] direction (Fig. 2a), reveals a repetitive pattern of a single wide terrace (with uniform straight edges) followed by five shorter terraces (more irregular edges). This pattern continuously repeats within this region of the sample, where the wide terrace comprises one surface domain and the five shorter terraces are the other (highlighted in Fig. 2a with yellow dashed lines). The topography data also reveals that the single wide terrace has a much smoother surface morphology, while the terraces comprising the other domain appears to have numerous adatoms and vacancies giving it a rougher appearance (also see inset of Fig. 4b). It is known that $TiO_2$ forms a stable surface termination and is commonly prepared via chemical processing[23, 24], while theoretical calculations indicate that SrO is a higher energy surface with a low desorption energy indicating it is less stable[25]. It has also been shown experimentally that the SrO surface termination on a layered perovskite, which possess a cleavage plane, is in fact less stable and results in numerous vacancies and adatoms[26].

A height profile (Fig. 2b) across one of the repetitive series of step-edges that encompasses the domain boundaries (solid yellow line in Fig. 2a) shows the clearest evidence that the striped domains are comprised of two different surface terminations. Within the highly stepped region the terrace height correlates to the lattice constant of a unit cell, roughly 4 Å, except at the boundaries where the heights are always either a half unit cell (2 Å) or one and half unit cells (6 Å), which indicates a change in the surface termination. Combining these observations with the surface morphology, our initial interpretation of this region can be summarized in Fig. 2c, where a wide low-energy domain of $TiO_2$ is followed by five tightly bunched steps terminated with SrO comprising the other domain. These domains continuously repeat across the surface.

Investigating the topography in regions where the step-edges are not running perpendicular to the [100] direction (Fig. 3a) the alternating striped domain pattern is more visible. Rather than domain boundaries being defined by step-edges, the transition from one termination to the other is occurring on the terrace, as highlighted by yellow dashed lines in Fig. 3a. In fact, there are two relatively distinct directions that the step-edges run depending upon which domain they are currently in. For instance, the step-edges with SrO termination span the domain at a shallower angle (blue dashed line in Fig. 3a) compared the $TiO_2$ domain, in which the edges, on average, cut sharply across the domain (red dashed line in Fig. 3a). Taken with the observation of step-bunching in the SrO domain, it would appear that it is energetically favorable to consolidate the step edges in the SrO domain.

The *dI/dV* measurements taken over regions believed to be terminated with $TiO_2$ are plotted (red curves) in Fig. 3b. These curves consist of both forward and backward sweeps taken on two different $TiO_2$ domains. A relative peak in the empty states of this region is observed at roughly +2.25 V. Similar *dI/dV* measurements were made over SrO domains (black curves) and are also plotted in Fig. 3b. The empty state LDOS (positive sample bias) of the SrO regions begin to increase at a higher energy than the $TiO_2$ regions, which it quickly surpasses in magnitude. Previously reported theoretical calculations of the LDOS for d-band electrons have shown that the $TiO_2$ empty states occur at a lower energy than the SrO states[27, 28]. Therefore, our assignment of the two different domains is consistent with these results.

Spatial conductance maps were taken at energies of interest determined from the individual *dI/dV* spectrum. Figure 3c shows a topographic image taken in a region where the step-edges are perpendicular to the [100]. Conductance maps taken at energies of +3.0 V and +2.25 V are displayed below the topography. These energies were chosen to show the transition in the LDOS of the two domains. At the higher energy the SrO domains have a higher conductance, while at lower energy the LDOS of the $TiO_2$ domains are more prevalent resulting in a higher conductivity. The conductance maps, especially at +3.0 V, illustrate how uniform the electronic properties of these domains are.

On a different sample, larger scale STM images were acquired to illustrate the long-range uniformity of the striped pattern. Figure 4a shows a large area scan taken over a highly-stepped region, in which the terraces are not aligned perpendicular to the [100]. The stripe pattern is clearly visible within this topography data and as previously observed the stripe pattern itself does run perpendicular to the [100]. The striped pattern is easily observed in the topography because the SrO domains are all physically bumped up 2 Å from the neighbor $TiO_2$ domains.

This illustrates that even for rough fractured surfaces, alternating domains are still prevalent. On average, the characteristic length scale of 10 nm to 30 nm for the two domains is also preserved. When zooming in with the STM (inset Fig. 4a) one can see that the shallow and steep angles of the step edges within the two domains (highlighted by the blue and red dashed lines) are consistent with the previous observations on different samples. In contrast, recent efforts have successfully optimized the fracture technique, as illustrated in Fig. 4b showing large atomically flat terraces that can be reproducibly obtained. The inset of Fig. 4b, showing a zoomed in region, confirms that the stripe pattern is preserved and uniform over an extremely large area of the surface.

In summary, after successful implementation of cross-sectional STM to a 3-dimensional perovskite oxide we have uncovered the formation of striped domains of alternating surface terminations with characteristic widths that form when fracturing $SrTiO_3$ *in situ* at room temperature. With STM characterization and spectroscopy we were able to assign the SrO termination to the wider domains and the $TiO_2$ termination to the narrower domains. These domains exhibit long-range order over the surface and are possibly driven by electrostatic interactions resulting from the weak polar nature of the surface termination. This argument is motivated by recent observations reported when preparing vicinal $SrTiO_3$[29, 30]. The success of this proof-of-principle experiment will undoubtedly help to advance new techniques for studying oxide interfaces. Current efforts are underway to apply cross-sectional STM to investigate the interface between oxide thin films and superlattices.

ACKNOWLEDGMENT : The use of the Center for Nanoscale Materials at Argonne National Laboratory was supported by the U.S. Department of Energy, Office of Science, Office of Basic Energy Sciences, under Contract No. DE-AC02-06CH11357.

FIGURE CAPTIONS

**Figure 1.** (a) Schematic showing the 3-dimensional unit cell for SrTiO3, which does not have a natural cleavage plane. (b) The top gradient enhanced STM image shows a highly stepped region of atomically flat terraces consisting of {100} planes (tunneling conditions: +3.00 V, 100 pA). (c) The spatial conductance map overlayed onto the topographic rendering. A well defined stripe pattern, with alternating widths of characteristic length (12nm and 24 nm), is clearly observed and illustrates a strong contrast in the LDOS on the surface that exhibits long-range order.

**Figure 2.** (a) The two different striped domains are highlighted in the topography image by yellow dashed lines. In this region of the sample one domain is always a wide terrace, with well defined step-edges, followed by the other domain consisting of five shorter terraces with irregular edges. The wide terrace domain has a much smoother surface morphology than the other domain. (b) This height profile (solid yellow line in (a)) shows the strongest evidence of a change in surface termination. The central terraces for the highly-stepped domain are all consistent with the lattice constant of the unit cell (~ 4 Å), except at the boundaries where the height difference is either half a unit cell (~ 2 Å) or one and a half unit cells (~ 6 Å), which indicates a change in surface termination. (c) This model represents the observed surface structure illustrating the regions of $TiO_2$ and SrO termination.

**Figure 3.** (a) Two of the domains are highlighted by yellow dashed lines. However, in this region the step-edges are not running perpendicular to the [100]. The SrO termination results in step-edges that follow a shallow angle across the domain (blue dashed line), in contrast to the TiO$_2$ terminated step-edges that sharply cross their domain (red dashed line). (b) Individual *dI/dV* spectra were measured over the two different domains. These reproducible spectra taken over the two domains indicate that the LDOS for TiO$_2$ is higher than SrO for lower energies but reverses at higher energies. (c) To illustrate the transition in conductivity for the two domains, spatial conductance maps were concurrently measured with the topography data illustrated at the top. The two spatial maps taken at +3.0 V and +2.25 V clearly show the transition and were chosen based on the spectra measured in (b).

**Figure 4.** (a) This large-scale STM image illustrates the long-range order of the striped domains. The striped domains are clearly visible in the topography because the SrO regions are all bumped up ~ 2 Å. The inset shows a zoom in of this area where the angles of the step-edges (blue and red dashed lines) are spanning the domains (yellow dashed lines) at angles that are consistent with previously fractured samples. (b) This STM image shows that after optimizing the fracture procedure we can reproducibly prepare surfaces with large terraces that are atomically flat. The inset shows that striped alternating surface termination is still preserved and extends over a large area of the surface.

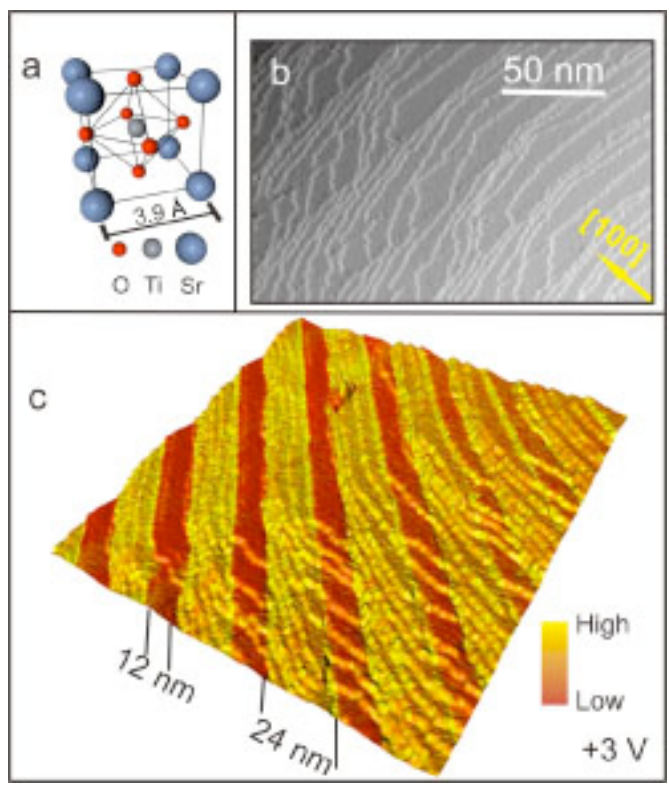

Figure 1, N. P. Guisinger *et al.*

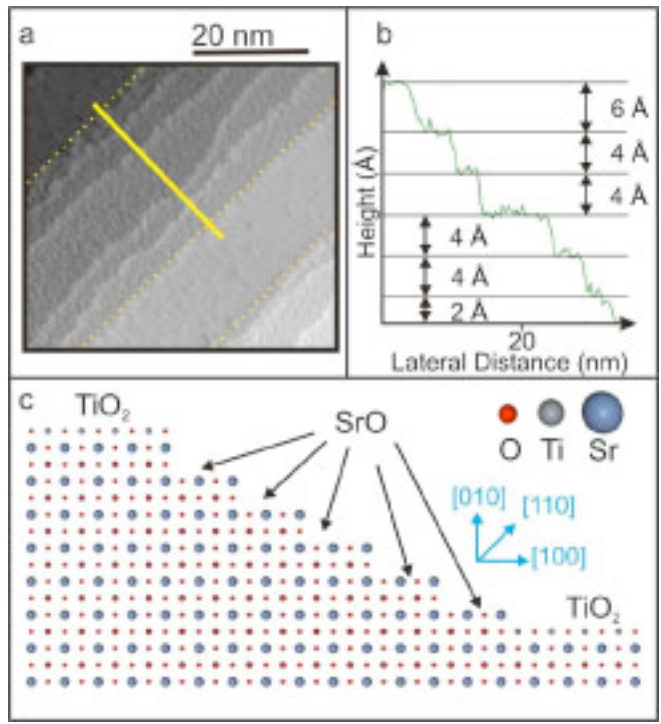

Figure 2, N. P. Guisinger *et al.*

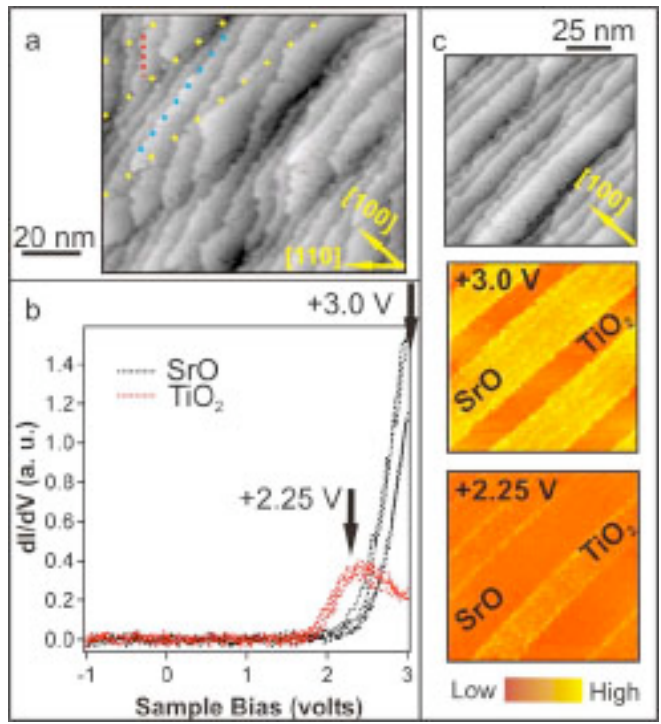

Figure 3, N. P. Guisinger *et al.*

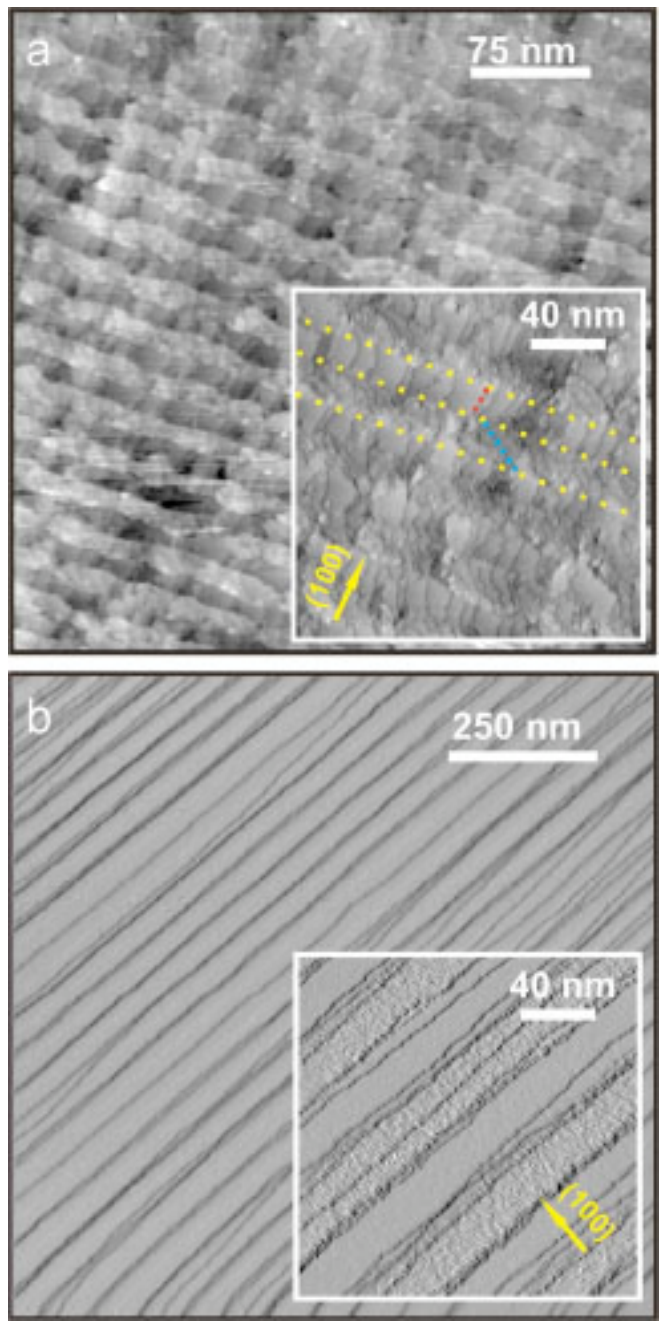

Figure 4, N. P. Guisinger *et al.*